\documentclass[%
reprint,
superscriptaddress,
 amsmath,amssymb,
pra,
]{revtex4-2}

\usepackage{graphicx}
\usepackage{dcolumn}
\usepackage{bm}
\usepackage{braket}

\begin{document}


\title{Electrically detected magnetic resonance of $^{75}$As magnetic clock transitions in silicon}

\author{Ravi Acharya}
\thanks{These authors contributed equally to this work.}
\affiliation{%
School of Physics, University of Melbourne, Parkville VIC 3010, Australia
}%
\affiliation{%
Photon Science Institute, Department of Electrical and Electronic Engineering, University of Manchester, Manchester, M13 9PL, United Kingdom
}%
\affiliation{%
London Centre for Nanotechnology, University College London, 17-19 Gordon Street, London, WC1H 0AH, United Kingdom
}%

\author{Shao Qi Lim}
\thanks{These authors contributed equally to this work.}
\affiliation{%
School of Physics, University of Melbourne, Parkville VIC 3010, Australia
}%
\affiliation{%
Australian Research Council Centre for Quantum Computation and Communication Technology}%
\affiliation{%
Department of Physics, School of Science, RMIT University, Melbourne 3001 VIC Australia.
}%

\author{Brett C. Johnson}  
\affiliation{%
Department of Physics, School of Science, RMIT University, Melbourne 3001 VIC Australia.
}%

\author{Nicholas Gillespie}
\affiliation{%
School of Physics, University of Melbourne, Parkville VIC 3010, Australia
}%
\affiliation{%
Australian Research Council Centre for Quantum Computation and Communication Technology}%
\affiliation{%
Department of Physics, School of Science, RMIT University, Melbourne 3001 VIC Australia.
}%

\author{Christopher T.-K. Lew}
\affiliation{%
Department of Physics, School of Science, RMIT University, Melbourne 3001 VIC Australia.
}%

\author{Alexander M. Jakob}
\affiliation{%
School of Physics, University of Melbourne, Parkville VIC 3010, Australia
}%
\affiliation{%
Australian Research Council Centre for Quantum Computation and Communication Technology}%

\author{Daniel L. Creedon}
\affiliation{%
CSIRO Manufacturing, Clayton, VIC 3168, Australia
}%

\author{Gus O. Bonin}
\affiliation{%
School of Physics, University of Melbourne, Parkville VIC 3010, Australia
}%

\author{Dane R. McCamey}
\affiliation{%
School of Physics, University of New South Wales, Sydney NSW 2052, Australia
}%

\author{Richard J. Curry}
\affiliation{%
Photon Science Institute, Department of Electrical and Electronic Engineering, University of Manchester, Manchester, M13 9PL, United Kingdom
}%

\author{Jeffrey C. McCallum}
\affiliation{%
School of Physics, University of Melbourne, Parkville VIC 3010, Australia
}%
\affiliation{%
Australian Research Council Centre for Quantum Computation and Communication Technology}%

\author{David N. Jamieson}
\email{d.jamieson@unimelb.edu.au}
\affiliation{%
School of Physics, University of Melbourne, Parkville VIC 3010, Australia
}%
\affiliation{%
Australian Research Council Centre for Quantum Computation and Communication Technology}%

\begin{abstract}
Magnetic clock transitions (CTs), defined by vanishing first-order sensitivity of the transition frequency to magnetic field fluctuations, provide a powerful route to suppress decoherence in donor spin systems. Here, we present the observation of magnetic field CTs from an ensemble of near-surface $^{75}$As ($I = 3/2$) spins in silicon using low-field ($< 10$~mT) continuous-wave electrically detected magnetic resonance (EDMR). As the CT condition is approached, pronounced linewidth broadening is observed, consistent with a donor Hamiltonian informed linewidth model. These results establish low-field EDMR as a sensitive probe of CTs in near-surface donor systems relevant to silicon-based quantum devices.

\end{abstract}

\maketitle


\section{Introduction}

Group-V donor spins in silicon constitute exceptional solid-state spin systems, combining long intrinsic coherence times and compatibility with well-established semiconductor fabrication processes~\cite{Tyryshkin2012,Muhonen2014,Zwanenburg2013}. Both the $^{31}$P donor ($I=1/2$) and heavier group-V species such as $^{75}$As ($I=3/2$), $^{121,123}$Sb ($I=5/2,\ 7/2$), and $^{209}$Bi ($I=9/2$) have been studied using electron spin resonance (ESR), nuclear magnetic resonance, and electrically detected magnetic resonance (EDMR) techniques~\cite{Stegner2006,Morishita2009,Franke2014,Franke2015,Mohammady2012,Zhu2017, Hori2021}. The larger nuclear spins of the heavier donors provide access to higher-dimensional Hilbert spaces of $N_{dim}=2I+1$. Compared with two-dimensional qubits, such qudits ($N_{dim}>2$) enable expanded control and encoding protocols that may offer advantages for scalable silicon-based quantum information processing architectures.\cite{Yu2025}

A central challenge for all donor spin species is the mitigation of decoherence arising from environmental noise. Considerable progress has been achieved through $^{28}$Si isotopic enrichment of the silicon host to suppress spectral diffusion from the residual $^{29}$Si nuclear spin bath~\cite{Acharya2024,Lim2025} as well as through dynamical decoupling techniques~\cite{Morton2008,Viola1998}. Complementary to these approaches, the donor spin Hamiltonian itself can be exploited by operating at optimal bias fields where the transition frequency is intrinsically insensitive to perturbation. These clock transitions (CTs) are defined by a vanishing first-order sensitivity to fluctuations in an external control parameter, such as the electric or magnetic field. In the case of magnetic CTs, $df/dB_0 = 0$, where $f$ is the spin transition frequency and $B_0$ is the applied magnetic field. For silicon donors, magnetic CTs often arise from strong electron–nuclear spin-mixing in the intermediate-field regime where the hyperfine interaction becomes comparable to the electron Zeeman energy ($A \sim\gamma_{e}B_0$ where $A$ is the isotropic hyperfine coupling constant and $\gamma_{e}$ the electron gyromagnetic ratio). Such transitions have been shown to substantially enhance coherence times in ensemble ESR measurements of $^{209}$Bi donors.\cite{Wolfowicz2013}\\

While CTs have been extensively investigated using ESR, their observation in EDMR, particularly in near-surface ensembles relevant to quantum devices, remains sparse. EDMR provides a sensitive means to probe donor spin transitions in a fully fabricated silicon device,\cite{Mccamey2006,Willems2008} particularly at low magnetic fields and in near-surface environments where conventional ESR can be limited.~\cite{Morishita2009,Dreher2015} In silicon devices, EDMR commonly relies on the spin-dependent recombination (SDR) mechanism between donor electrons and paramagnetic interface defects. In particular, the $P_{b0}$ centre, a silicon dangling-bond defect at the Si/SiO$_2$ interface,\cite{Stesmans1998} forms coupled spin pairs with nearby donors, such that the recombination rate depends on the spin symmetry of the pair.\\

In this work, we employ EDMR to investigate low-field magnetic CTs of near-surface $^{75}$As donors in silicon. Turning points in the transition spectrum corresponding to magnetic CTs are identified, and a pronounced increase in spectral linewidth is observed as the CT condition is approached. This broadening is described in terms of modulation-induced dispersion and variations in the magnetic and hyperfine coefficients of the device. 

\section{Experimental Methods}

\begin{figure}
\begin{center}
\rotatebox{0}{\includegraphics[width=\linewidth]{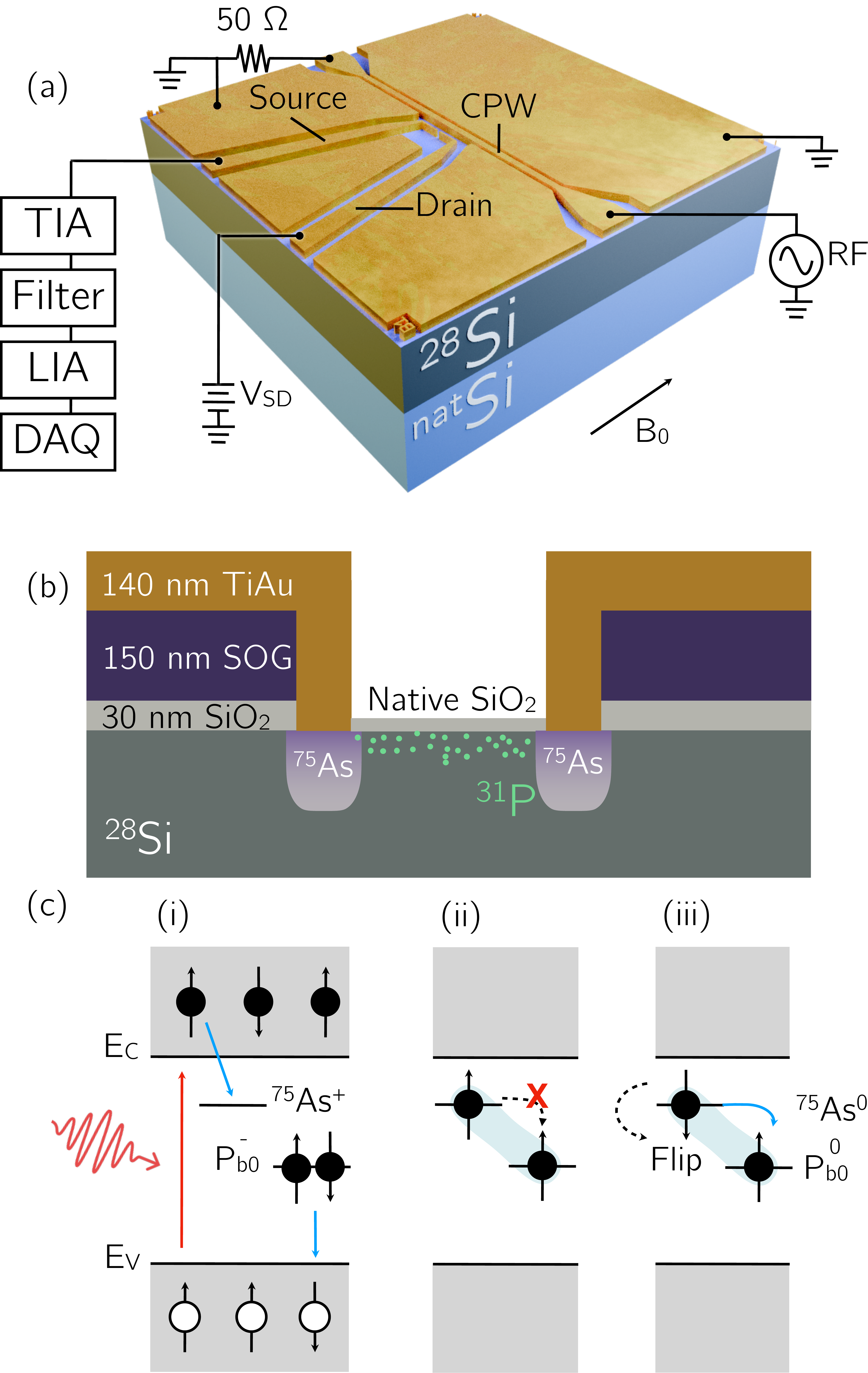}}
\end{center}
\caption[]{(a) EDMR device schematic, showing the coplanar waveguide (CPW), the source-drain contacts, and auxiliary electronics including: a transimpedance amplifier (TIA), a lock-in amplifier (LIA), a data acquisition system (DAQ), a DC-bias source, a RF source, and a 50 $\Omega$ terminator. 
(b) Cross-sectional view of the device geometry (not-to-scale), illustrating the near-surface donor layer. 
(c) Schematic of the SDR mechanism. (i) Above bandgap (633 nm) excitation generates electron–hole pairs, enabling the formation of weakly coupled donor–defect spin pairs. Recombination is spin selective: (ii) parallel pairs are long lived, while (iii) antiparallel pairs lead to recombination events. Resonant RF excitation redistributes the spin pair populations, modulating the recombination rate and hence the device conductivity.} 
\label{F1}
\end{figure}

\subsection{Device Fabrication}

A schematic of the EDMR device geometry is shown in Figs.~\ref{F1}(a) and (b). The device was fabricated on an isotopically enriched $^{28}$Si substrate (800~ppm residual $^{29}$Si), consisting of a 10-$\mu m$-thick epitaxial layer grown via chemical vapor deposition on a p-type natural silicon wafer (Isonics Corp.). A 30-nm-thick thermal oxide was initially grown and patterned to define selective in-diffusion areas. Source and drain electrodes were formed by the drive-in diffusion of $^{75}$As from a spin-on glass (SOG) source (Desert Silicon) at 950 $^\circ$C for 10~minutes in an ArH ambient.

Following the removal of the SOG and thermal oxide via hydrofluoric acid etching, the $40\times40\;\rm \mu m^2$ active region was implanted with 9~keV $^{31}$P ions at an $8^\circ$ tilt to a fluence of $10^{12}$~cm$^{-2}$. While the $^{31}$P donors were intended for the study of the depleted $^{29}$Si nuclear spin bath, the $^{75}$As donors investigated in this work originated from lateral diffusion from the source-drain contacts into the active area during subsequent thermal processing. This lateral diffusion was modelled to confirm the presence of $^{75}$As within the sensing volume (see Supplementary Fig.~S1 \cite{supplemental_material} which includes Refs. \cite{Ziegler2008} and \cite{Newman1983}). Lattice damage repair and activation of the P and As donors was achieved using a two-step rapid thermal anneal: 620 $^\circ$C for 10~minutes followed by 1000 $^\circ$C for 5 s in Ar. A native oxide layer formed over the active device region, resulting in $P_{b0}$ defects which act as deep level recombination centres as required for SDR.

For on-chip RF delivery, a coplanar waveguide (CPW) was patterned on the substrate with a geometry designed for broadband operation. The metallization consisted of a Ti/Au (20/120~nm) bilayer deposited via electron-beam evaporation. The completed device was wire-bonded to a printed circuit board (PCB) for integration into the EDMR measurement cryostat.

\subsection{EDMR Spectroscopy}

In this work, we are most interested in probing SDR processes that involve near-surface $^{75}$As donor electrons weakly coupled to $P_{b0}$ defects. In this section, we will first describe our EDMR setup followed by the $^{75}$As-$P_{b0}$ spin pair's SDR mechanism. 

EDMR was performed using a low-field spectrometer integrated with a Janis optical cryostat (operated at $T=5$ K). A schematic of key components of our experimental setup is shown in Fig.~\ref{F1}(a). The device was DC-biased at 500~mV through the input connector of a Stanford Research Systems SR570 low-noise transimpedance amplifier (TIA). To facilitate SDR, the active area was illuminated with a continuous-wave 633~nm red laser (TUI Optics DL100), generating a steady-state photocurrent of approximately 500~nA between the source and drain electrodes. The RF drive was provided by an Agilent E8267C signal generator, which was frequency modulated with a 1.737~kHz sine wave using the internal reference of a Stanford Research Systems SRS 830 digital lock-in amplifier (LIA). The source-drain current was converted to a voltage using the TIA and further conditioned using a 6~dB, 1–3~kHz bandpass filter and a fifth-order 600~Hz high-pass Chebyshev filter to optimize the signal-to-noise ratio. The voltage output was then demodulated with the LIA referenced to the applied RF modulation and subsequently collected by a National Instruments USB-6003 data acquisition system (DAQ). The static magnetic field ($B_0$) was generated by a pair of Helmholtz coils driven by a Kepco BOP 20-10M power supply. The $B_0$ field sweeps were controlled by a Rigol DG4162 waveform generator that provided a linear ramp to the power supply.

The SDR mechanism operative in our device is illustrated in Fig.~\ref{F1}(c) and was adapted from Ref.~\citenum{Hoehne2013}. In the absence of above-bandgap optical excitation, $^{75}$As donors are ionized ($^{75}$As$^+$) due to electron transfer to nearby $P_{b0}$ centres, which become negatively charged ($P_{b0}^-$), as depicted in Fig.~\ref{F1}(c)(i). Continuous-wave 633~nm illumination generates a steady population of electron-hole pairs that neutralize these charged states via carrier capture and recombination (blue arrows), enabling formation of weakly coupled spin pairs, each with $S=1/2$. 

In the weak exchange coupling limit, the spin pair eigenstates are either parallel (Fig.~\ref{F1}(c)(ii)) or antiparallel (Fig.~\ref{F1}(c)(iii)) spin pairs, $\ket{\uparrow\uparrow}, \ket{\downarrow\downarrow}, \ket{\uparrow\downarrow}, \ket{\downarrow\uparrow}$. \cite{Boehme2003,Hoehne2013} The antiparallel states undergo rapid recombination ($\tau \sim 10~\mu$s) while the parallel states are blocked by the Pauli exclusion principle and are therefore long lived ($\tau \sim 1$ ms).~\cite{Hoehne2013} As a result, the steady-state population is predominately parallel spin pairs (Fig.~\ref{F1}(c)(ii)). 

When the applied RF field is resonant with the $^{75}$As transition, it drives electron spin flips within the donor–defect pair (dashed arrow), converting a fraction of the long-lived parallel pairs into antiparallel configurations (Fig.~\ref{F1}(c)(iii)). The ensuing rapid recombination of these antiparallel pairs reduces the steady-state photo-carrier density, producing a measurable decrease in the source–drain current.

\section{Results and discussion}

\begin{figure*}[t]
\begin{center}
\rotatebox{0}{\includegraphics[width=14cm]{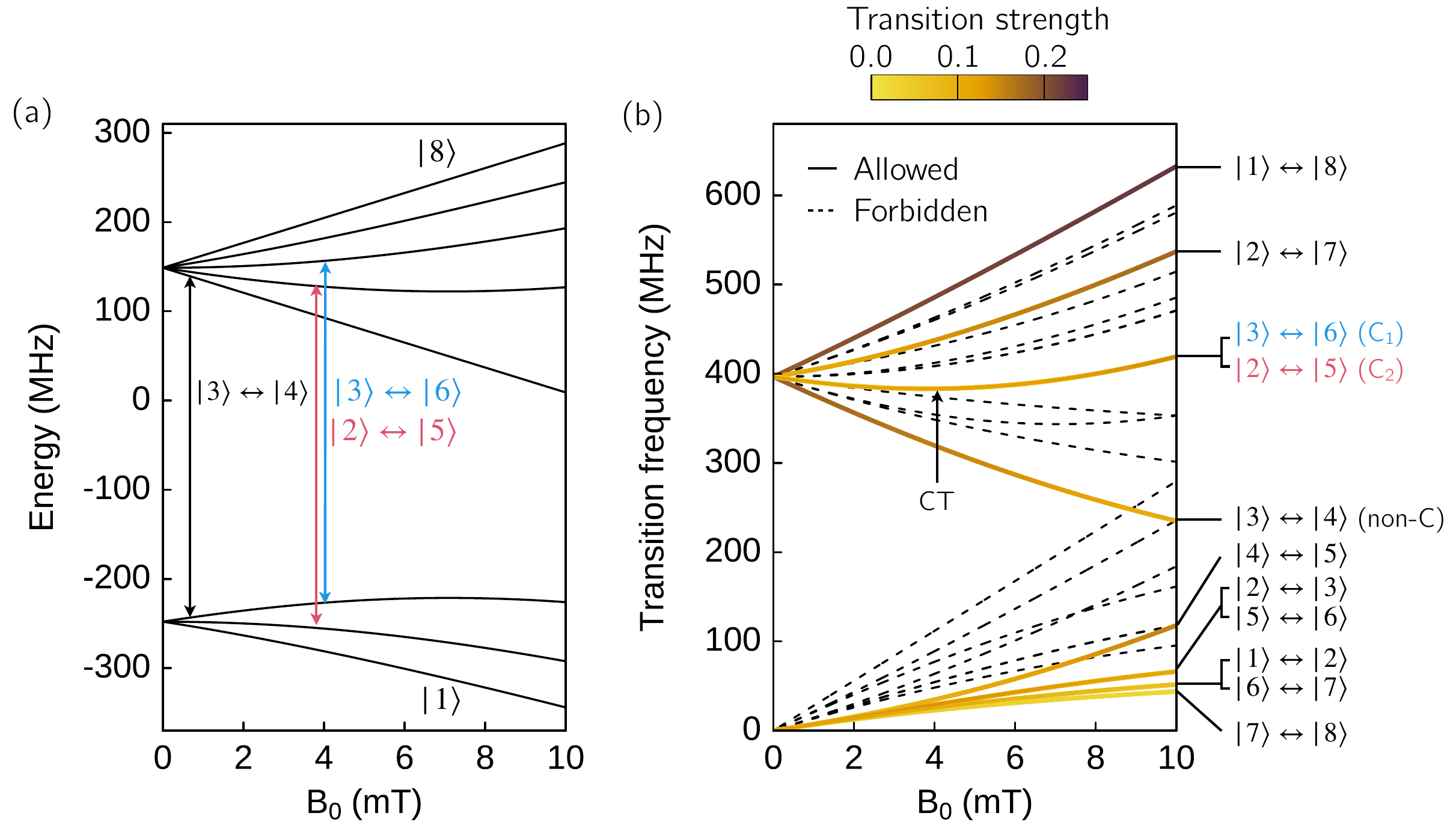}}
\end{center}
\caption[]{(a) Calculated energy level diagram of substitutional $^{75}$As in silicon as a function of the applied magnetic field $B_0$, labelled as $\ket{1}-\ket{8}$ in ascending order of energy. The two CTs are indicated by the red and blue vertical lines. The black vertical line indicates the transition labelled non-C in (b). (b) Corresponding transition frequencies between the donor eigenstates. Both the magnetically allowed transitions (solid lines) and forbidden transitions (dashed lines) are shown. The colour scale of the allowed transitions encodes the corresponding ESR transition strengths. Each transition is labelled by its associated eigenstate indices on the right. The CT transitions (C$_1$ and C$_2$, designated the CT pair in the text) appear as turning points where $df/dB_0=0$ (indicated by the black arrow). The transition labelled non-C has a role described in the text.} 
\label{F2}
\end{figure*}

The resonant energies of the magnetic resonance signals were calculated using the Hamiltonian for the electron-nuclear spin system of substitutional $^{75}$As donors in silicon, 

\begin{equation}
\hat{H} = g_e \mu_B B_0 \hat{S}_z - g_n \mu_N B_0 \hat{I}_z + A\, \hat{\mathbf{S}} \cdot \hat{\mathbf{I}},
\label{eq:Hamiltonian}
\end{equation}

\noindent where $A=198.4$~MHz is the isotropic hyperfine coupling constant for $^{75}$As,\cite{Feher1959} and  $\hat{\mathbf{S}}$ and $\hat{\mathbf{I}}$ are the electron ($S=1/2$) and nuclear ($I=3/2$) spin operators, respectively. $\hat{S_z}$ and $\hat{I_z}$ are the electron and nuclear spin operators along the $z$-projection axis, $B_0$ is the magnitude of the applied magnetic field, $\mu_B$ and $\mu_N$ are the Bohr and nuclear magnetons, and $g_e=1.99837$ and $g_n=0.959$ are the electron and nuclear $g$-factors.\cite{Feher1959} The spin $z$-projection axis is defined to be parallel to the applied $B_0$ field. Although we use the SDR mechanism for readout of the $^{75}$As electron spin, the simplified Hamiltonian above does not include the $^{75}$As-P$_{b0}$ spin-pair interactions, due to the weakness of the spin-pair coupling.

The interplay between the electron and nuclear Zeeman interactions and the isotropic hyperfine coupling gives rise to strongly mixed eigenstates at low-to-intermediate magnetic field strengths (where $(\gamma eB_0<<A)$ to $(\gamma e B_0\sim A)$ which is $\sim$7~mT). Diagonalization of Eq.~\ref{eq:Hamiltonian} yields eight such eigenstates and their energy dependence as a function of $B_0$ is shown in Fig.~\ref{F2}(a). Allowed and forbidden magnetic dipole transitions between eigenstates $|i\rangle$ and $|j\rangle$ are characterized by their transition frequencies, $f$, and transition strengths, $S_{ij} = \left| \langle i | \hat{S}_x | j \rangle \right|^2$, where $\hat{S_x}$ is the electron spin operator along the $x$-projection axis. These are plotted in Fig.~\ref{F2}(b). 

Magnetic CTs occur at turning points of a transition, where $df/dB_0 = 0$. In $^{75}$As, there are two closely spaced magnetic CTs both located around 3.8~mT with a resonant excitation frequency of $\sim$ 383~MHz. These are associated with the minima of the $\ket{2}\leftrightarrow\ket{5}$ and $\ket{3}\leftrightarrow\ket{6}$ transitions, which are labelled C$_1$ and C$_2$, respectively (see Figs.~\ref{F2}(a) and (b)). C$_1$ and C$_2$ are hereafter designated the CT pair. At these operating points, spin decoherence arising from magnetic field noise is suppressed to first order. An additional transition of note is $\ket{3}\leftrightarrow\ket{4}$, referred to here as non-C (a non-clock $^{75}$As transition), which appears at lower fields around 0.65~mT at 383~MHz and will be compared to C$_1$ and C$_2$ below. 

\begin{figure}
\begin{center}
\rotatebox{0}{\includegraphics[width=\linewidth]{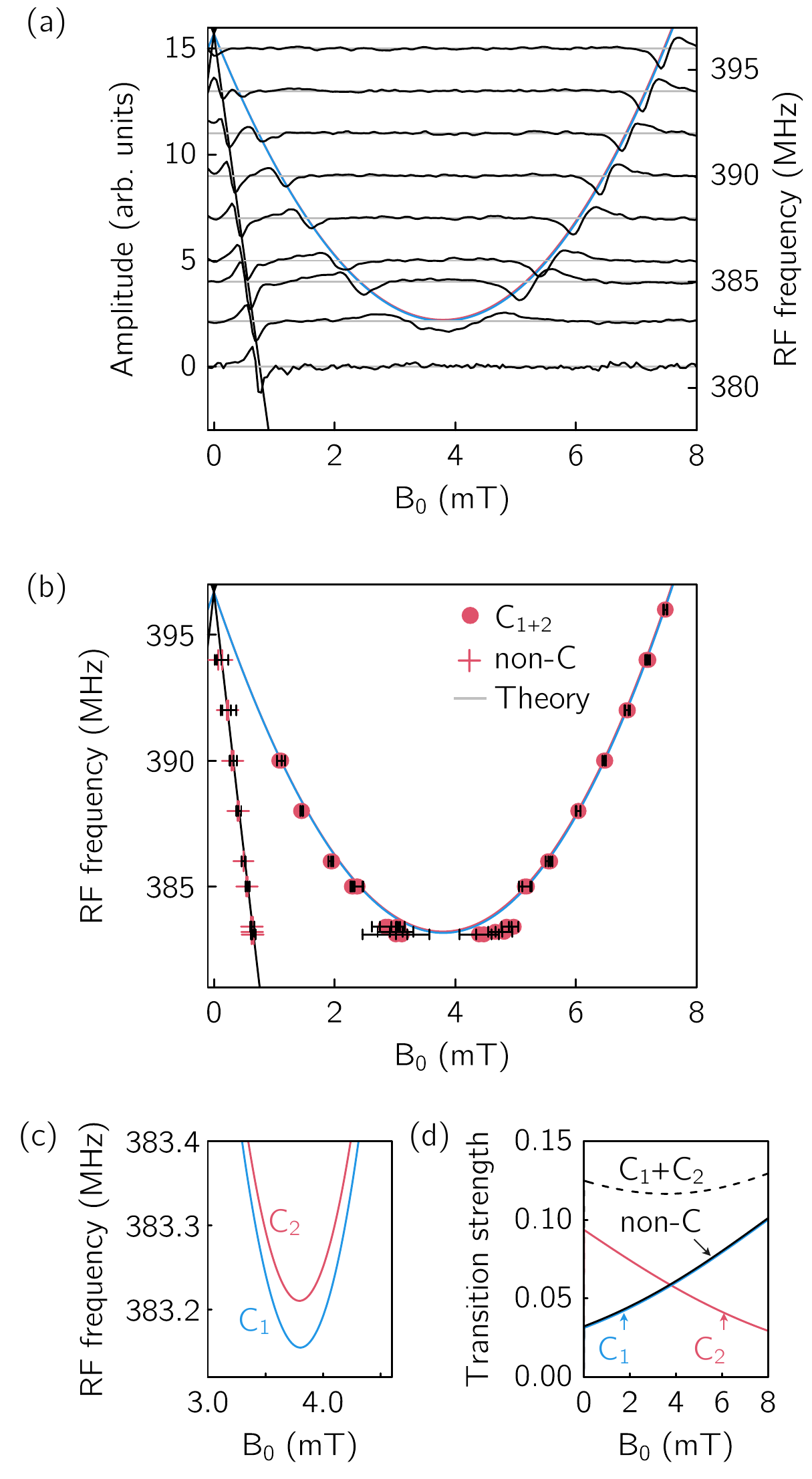}}
\end{center}
\caption[]{(a) Series of EDMR spectra versus $B_0$ with the right y-axis denoting the RF frequencies used (grey horizontal lines). The theoretically calculated transition frequency spectra are also shown (curved solid red/blue lines and straight black line). (b) Comparison between the experimentally measured (data points) extracted from fits to the data in (a) and theoretically calculated (solid lines) transition frequency spectra near the CT turning point. The solid black line and crosses represent the non-C transition while the solid red/blue lines and circles represent the CT pair. (c) Close-up near the CT turning point, detailing the small energetic and magnetic field separation between the CT pair. (d) Calculated ESR transition strengths versus $B_0$.} 
\label{F3}
\end{figure}

Fig.~\ref{F3}(a) shows a series of $^{75}$As EDMR spectra as a function of $B_0$ near the CT turning point. Spectra were collected over several magnetic field sweeps until a sufficient signal-to-noise ratio was obtained. These were collected with a microwave power of 19~dBm and a RF frequency modulation (FM) amplitude of 500~kHz. A spline fit was used to subtract the non-resonant background from the raw spectra. At lower RF frequencies the $P_{b0}$ centre can also be observed in the same sample, under the same conditions (see Supplementary Fig.~S4 \cite{supplemental_material}). Each spectrum in the frequency range considered here consists of three distinguishable resonance features: the non-C transition is represented by the resonance feature in the low-field region of the spectra while the C$_1$ and C$_2$ transitions are represented by the other two resonance features (even though four are expected, as will be discussed later in relation to Fig.~\ref{F3}(c)). We note that the high-field resonance appears inverted in phase. This is due to the RF-FM lock-in technique employed here where $\Delta f_{\mathrm{mod}}$, the RF-FM amplitude,  is mapped into an effective magnetic field modulation according to:

\begin{equation}
\label{eq:mod}
    \Delta B_{\mathrm{mod}} = \frac{dB_0}{df}\,\Delta f_{\mathrm{mod}}.
\end{equation}

\noindent Consequently, the sign (or phase) of the detected lock-in signal depends on the slope, $df/dB_0$. This is well-reproduced by simulations (see Supplementary Fig.~S2 \cite{supplemental_material}). At low-field strengths before the CT turning point, the slope is negative, whereas beyond this turning point (towards higher fields) it becomes positive, resulting in a phase inversion of the lock-in signal across the CT turning point.

Each resonance in the spectra was fitted with a single phenomenological skewed Lorentzian to extract the ${}^{75}$As resonance positions and peak-to-peak linewidth, and also to deconvolve features that consisted of multiple superimposed resonances. These values are compared with the calculated transition frequencies in Fig.~\ref{F3}(b) and show good agreement. A slight discrepancy appears near the theoretical turning point of the CT pair (3.8~mT, 383~MHz), which we attribute to challenges in fitting these features and indicated by the large error bars. 

A close up view of the C$_1$ and C$_2$ resonances at the turning points are shown in Fig.~\ref{F3}(c). The minima to C$_1$ and C$_2$ both occur at around 3.8~mT with a small 8~$\mu$T field offset. The frequency separation between C$_1$ and C$_2$ is 56~kHz, which is much smaller than the FM amplitude employed (500~kHz). The relatively small separations in both the resonant $B_0$ and $f$ meant the individual C$_1$ and C$_2$ resonances could not be resolved with the present apparatus. Instead the C$_1$ and C$_2$ pair contribute to a single resonance, with a slight linewidth broadening from the overlapping pair of resonances. We note that at $B_0=3.8$ mT, the ESR transition strengths of C$_1$ and C$_2$ are approximately equal as shown in Fig.~\ref{F3}(d). However, towards higher fields the EDMR resonance is expected to take on more of the C$_1$ character while at lower fields C$_2$ dominates. With improvements in detection electronics and a smaller FM amplitude, it may be possible to yield a better field resolution while retaining a reasonable signal-to-noise ratio.  

\begin{figure}
\begin{center}
\rotatebox{0}{\includegraphics[width=\linewidth]{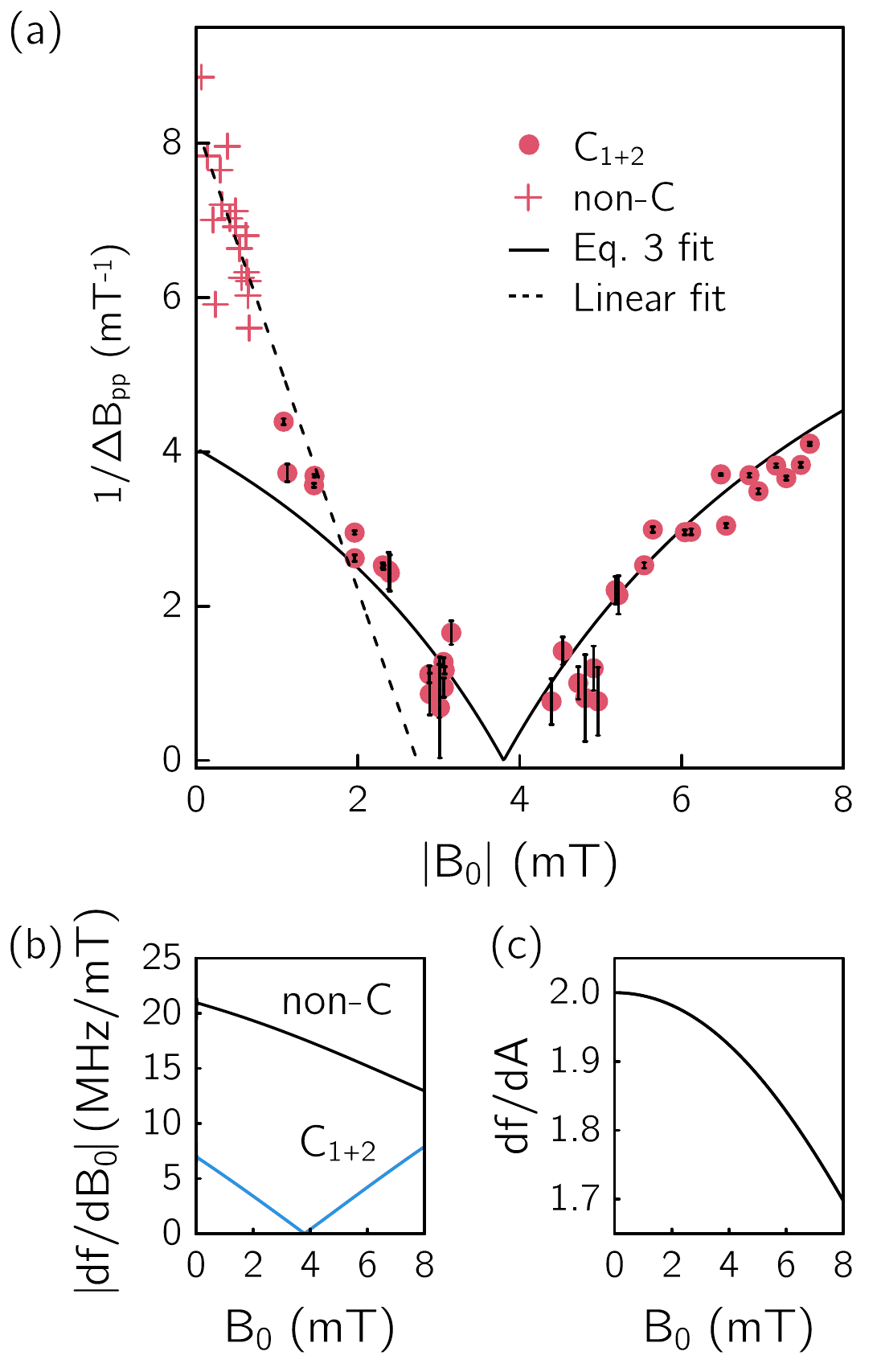}}
\end{center}
\caption[]{(a) Inverse of the peak-to-peak linewidth, $1/\Delta B_{\mathrm{pp}}$, extracted from Lorentzian fits to the EDMR spectra. The dashed grey line represents a linear fit to the non-C transition (crosses). The solid line is a fit using Eq.~\ref{eq:linewidth} to the CT pair transition (circles). (b) Absolute value of the derivative of the transition frequencies, $|\ df/dB_0\ |$, against $B_0$ for the relevant transitions. The CT pair (shown as C$_{1+2}$) is identified by the zero-crossings. (c) $df/dA$ against $B_0$ of the CT pair calculated from Eq.~\ref{eq:dfdA}.} 
\label{F4}
\end{figure}

Fig.~\ref{F4}(a) shows the inverse of the peak-to-peak linewidth, $1/\Delta B_{\mathrm{pp}}$, extracted from Lorentzian fits to the resonances presented in Fig.~\ref{F3}. Data are shown for the non-C and the CT pair transitions. A pronounced reduction in $1/\Delta B_{\mathrm{pp}}$ (corresponding to linewidth broadening) is observed for the CT pair resonance as the CT condition is approached.

To quantify this behaviour (to first-order), we express the peak-to-peak linewidth with the magnetic ($\Delta B_0$) and hyperfine ($\Delta A$) terms of the transitions,\cite{Wolfowicz2013}


\begin{equation}
\Delta B_{\mathrm{pp}} = \left( \Delta B_0 \right) +
\left( \frac{dB_0}{df} \right) \left( \frac{df}{dA} \Delta A \right) +
\left( \Delta B_{\mathrm{mod}} \right),
\label{eq:linewidth}
\end{equation}

\noindent where $\Delta B_0$ and $\Delta A$ parameterize the static inhomogeneities in the local magnetic field and hyperfine interaction, respectively.

The $\Delta B_0$ term may arise from: spatial variations in the applied field, dipolar fields from neighbouring paramagnetic centres, the $^{29}$Si spin bath, residual magnetic disorder at the Si/SiO$_2$ interface, or magnetic field gradients across the device. Such contributions are well-described within standard inhomogeneous broadening models for ESR. \cite{Poole1979} The hyperfine broadening, $\Delta A$, relates to local variations in the donor environment, including strain-induced variations of the hyperfine interaction, electric-field-induced Stark shifts, and microscopic disorder associated with implantation damage or proximity to interface defects.

The final term, $\Delta B_{\mathrm{mod}}$, in Eq.~\ref{eq:linewidth} accounts for effective broadening introduced by RF-FM as discussed earlier in relation to Eq.~\ref{eq:mod}. Eq.~\ref{eq:linewidth} therefore captures the mapping of frequency dispersion into field broadening as the magnetic field sensitivity is reduced. Both $\Delta B_{\mathrm{mod}}$ and the hyperfine term contain $df/dB_0$ which is plotted in Fig.~\ref{F4}(b).

The hyperfine term is calculated directly from the Hamiltonian eigenstates as,

\begin{equation}
\frac{df}{dA} =
\left\langle j \middle| \hat{\mathbf{S}}\cdot\hat{\mathbf{I}} \middle| j \right\rangle
-
\left\langle i \middle| \hat{\mathbf{S}}\cdot\hat{\mathbf{I}} \middle| i \right\rangle,
\label{eq:dfdA}
\end{equation}

\noindent where $\ket{i}$ and $\ket{j}$ denote the initial and final states of the transition. This term for the CT pair is plotted in Fig.~\ref{F4}(c).

The solid line in Fig.~\ref{F4}(a) is a fit to the CT pair transition data (solid red circles) using Eq.~\ref{eq:linewidth}. This yields values of $\Delta A = (0.26 \pm 0.05)$~MHz and $\Delta B_0 = (0.10 \pm 0.02)$~mT. Assuming that the strain modifies the hyperfine constant approximately linearly,\cite{Wilson1961} such that $\Delta A / A \sim \kappa \epsilon$, the extracted $\Delta A$ corresponds to a strain magnitude of $\epsilon \sim 7\times 10^{-5}$ with $\kappa=19.1$.\cite{Mansir2018} This level of strain is consistent with the proximity of the Si/SiO$_2$ interface and the high ${}^{75}$As concentration in the source/drain contact region. The parameter $\Delta B_0$ sets the minimum achievable linewidth away from the CT condition  and likely reflects  a combination of residual magnetic disorder in the device environment as well as instrumental broadening.


The dashed line in Fig.~\ref{F4}(a) is a linear fit, $mB_0+C$, to the non-C transition data (red crosses in Fig.~\ref{F4}(a)). The best fit is obtained with $m=-3.01\pm 0.7$~mT$^{-2}$ and $c=8.2 \pm 0.3$~mT$^{-1}$, corresponding to an offset linewidth of $\Delta B_{pp}\approx 0.12$~mT. While the linewidth model of Eq.~\ref{eq:linewidth} captures the qualitative increase in broadening with $B_0$ for this resonance (especially through the $df/dB_0$ term in Fig.~\ref{F4}(b)), it does not fully reproduce the magnitude of the observed dependence. Possible origins of this difference include additional magnetic disorder or unresolved spectral contributions and are discussed further in the supplemental material \cite{supplemental_material}.


At low magnetic fields ($B_0<2$~mT), deviations between experiment and calculation are observed for the CT pair transition. We attribute this to strong spectral convolution between nearby resonances, as evident in the spectra shown in Fig.~\ref{F3}(a). The resulting overlap reduces the accuracy with which individual linewidths can be extracted and likely leads to an underestimation of $\Delta B_{pp}$ in this magnetic field range.

The observed behaviour reflects the defining property of magnetic CTs, namely the reduced sensitivity of the transition frequency to fluctuations in the external magnetic field ($df/d B_0 \rightarrow 0$). As a consequence, magnetic disorder contributes less to the intrinsic frequency broadening near the CT condition. However, when the resonance is measured in the magnetic field domain, this reduced slope leads to an apparent divergence of the linewidth through the inverse factor $(dB_0/df)$ in Eq.~\ref{eq:linewidth}. The broadening observed near the CT condition therefore reflects the increasing influence of hyperfine disorder and the conversion between frequency and magnetic field units, rather than a loss of coherence of the underlying spin transition. Further improvements in signal-to-noise ratio may provide additional insights into the different linewidth broadening mechanisms operative within donor-$P_{b0}$ spin pairs.



\section{Conclusion}

We observe low-field magnetic CTs of near-surface $^{75}$As donor spins in silicon using continuous-wave EDMR. Detection is mediated by SDR between the donor electron and a proximal $P_{b0}$ interface defect, enabling sensitive electrical  readout. As the CT condition is approached, the EDMR linewidth is observed to increase markedly. This behaviour is well described by a model incorporating magnetic and hyperfine terms of the transition, together with modulation-induced broadening inherent to RF-FM lock-in detection. 

These results establish EDMR as a sensitive probe of magnetic CT physics in near-surface donor systems, directly relevant to silicon-based quantum device architectures. The ability to access CTs at low magnetic fields within a realistic interface environment comprised of a native oxide layer, provides a pathway to integrating field-insensitive donor spin transitions with electrically addressable device structures. The combination of Hamiltonian-level modelling and EDMR spectroscopy presented here provides a framework for engineering and identifying protected operating points in donor spin-based quantum technologies.

\begin{acknowledgments}
This work was funded by the Australian Research Council Discovery Project DP220103467 and the University of Melbourne/ University of Manchester collaboration scheme. D.N.J acknowledges the support of a Royal Society (UK) Wolfson Visiting Fellowship RSWVF/211016. R.A acknowledges the support of a Melbourne Research Scholarship. We thank Thomas Schenkel for providing the Isonics $^{28}$Si material used in this study that was bought with funds from NSA/LPS/ARO in the mid 2000s. R.J.C acknowledges the support of EPSRC grants EP/R025576/1, EP/V001914/1 and EP/R00661X/1 (Henry Royce Institute) and capital investment by the University of Manchester. We acknowledge the National Collaborative Research Infrastructure Scheme Heavy Ion Accelerator Platform for providing access to ion implantation facilities.
\end{acknowledgments}

\bibliography{bib}%

\end{document}


\title{Supplemental material:\\
Electrically detected magnetic resonance of $^{75}$As magnetic clock transitions in silicon}

\author{Ravi Acharya}
\thanks{These authors contributed equally to this work.}
\affiliation{%
School of Physics, University of Melbourne, Parkville VIC 3010, Australia
}%
\affiliation{%
Photon Science Institute, Department of Electrical and Electronic Engineering, University of Manchester, Manchester, M13 9PL, United Kingdom
}%
\affiliation{%
London Centre for Nanotechnology, University College London, 17-19 Gordon Street, London, WC1H 0AH, United Kingdom
}%

\author{Shao Qi Lim}
\thanks{These authors contributed equally to this work.}
\affiliation{%
School of Physics, University of Melbourne, Parkville VIC 3010, Australia
}%
\affiliation{%
Australian Research Council Centre for Quantum Computation and Communication Technology}%
\affiliation{%
Department of Physics, School of Science, RMIT University, Melbourne 3001 VIC Australia.
}%

\author{Brett C. Johnson} 
\affiliation{%
Department of Physics, School of Science, RMIT University, Melbourne 3001 VIC Australia.
}%

\author{Nicholas Gillespie}
\affiliation{%
School of Physics, University of Melbourne, Parkville VIC 3010, Australia
}%
\affiliation{%
Australian Research Council Centre for Quantum Computation and Communication Technology}%
\affiliation{%
Department of Physics, School of Science, RMIT University, Melbourne 3001 VIC Australia.
}%

\author{Christopher T.-K. Lew}
\affiliation{%
Department of Physics, School of Science, RMIT University, Melbourne 3001 VIC Australia.
}%

\author{Alexander M. Jakob}
\affiliation{%
School of Physics, University of Melbourne, Parkville VIC 3010, Australia
}%
\affiliation{%
Australian Research Council Centre for Quantum Computation and Communication Technology}%

\author{Daniel L. Creedon}
\affiliation{%
CSIRO Manufacturing, Clayton, VIC 3168, Australia
}%

\author{Gus O. Bonin}
\affiliation{%
School of Physics, University of Melbourne, Parkville VIC 3010, Australia
}%

\author{Dane R. McCamey}
\affiliation{%
School of Physics, University of New South Wales, Sydney NSW 2052, Australia
}%

\author{Richard J. Curry}
\affiliation{%
Photon Science Institute, Department of Electrical and Electronic Engineering, University of Manchester, Manchester, M13 9PL, United Kingdom
}%

\author{Jeffrey C. McCallum}
\affiliation{%
School of Physics, University of Melbourne, Parkville VIC 3010, Australia
}%
\affiliation{%
Australian Research Council Centre for Quantum Computation and Communication Technology}%

\author{David N. Jamieson}
\email{d.jamieson@unimelb.edu.au}
\affiliation{%
School of Physics, University of Melbourne, Parkville VIC 3010, Australia
}%
\affiliation{%
Australian Research Council Centre for Quantum Computation and Communication Technology}%

\maketitle
\newpage

\section{Arsenic diffusion simulations}

\begin{figure}[h]
\begin{center}
\rotatebox{0}{\includegraphics[width=0.6\linewidth]{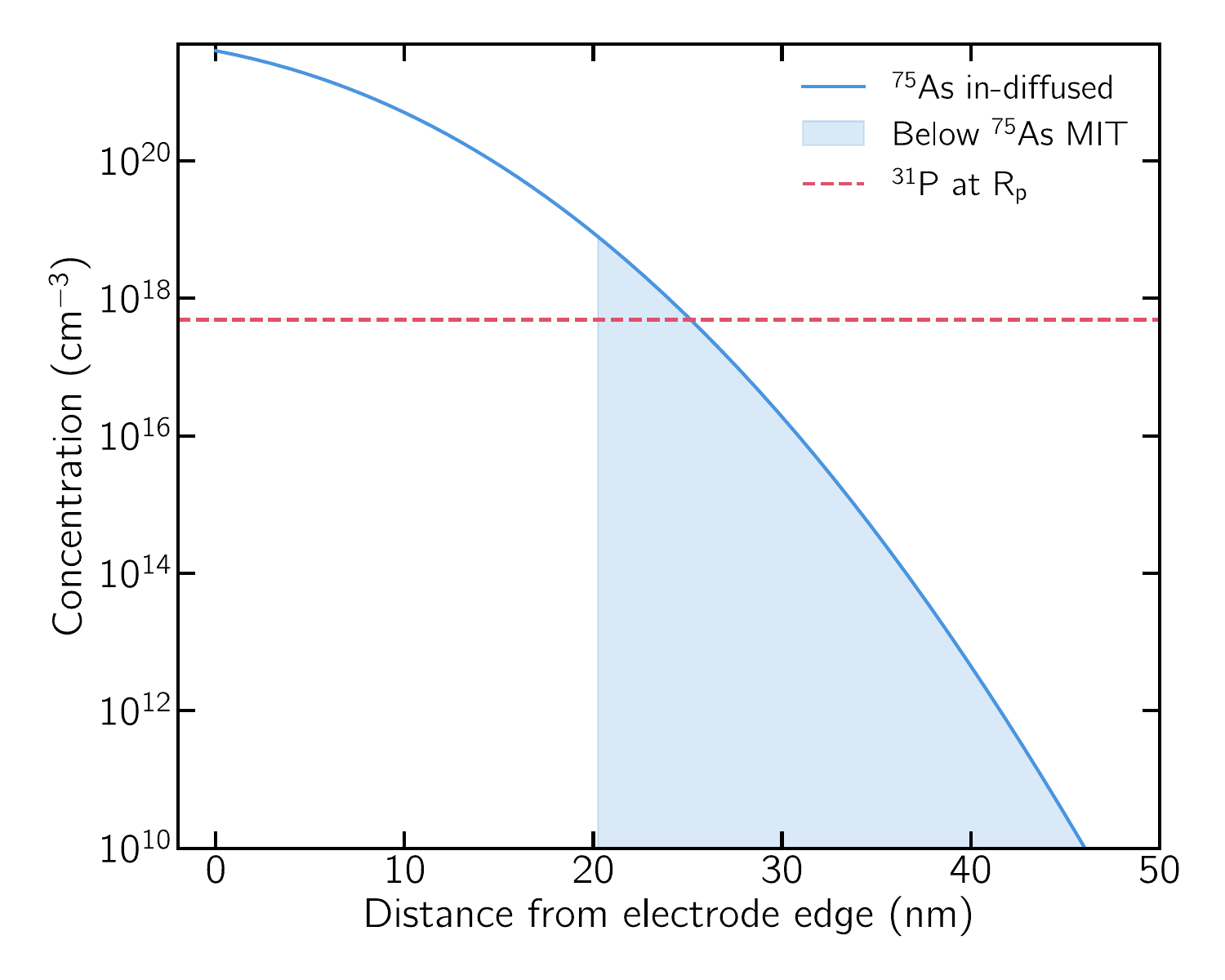}}
\end{center}
\caption[]{Lateral $^{75}$As diffusion profile from the edge of the source or drain contact. The blue shaded region denotes the area below the $^{75}$As metal-insulator transition where the $^{75}$As EDMR signal likely originates. The dotted red line shows the implanted $^{31}$P concentration at the ion-implant range ($R_p$) of $\sim 16$~nm.} 
\label{S1}
\end{figure}

Fig.~\ref{S1} shows lateral diffusion simulations of $^{75}$As using a simple Fickian diffusion model. The diffusion accounts for the full thermal budget employed during device fabrication which consisted of: a drive-in anneal at 950 $^\circ$C for 10~minutes under ArH, a rapid thermal anneal at 620 $^\circ$C for 10~minutes followed by 1000 $^\circ$C for 5~s in Ar.

The simulation is plotted as a function of distance from the edge of the electrode. The 9~keV $^{31}$P implant resulted in a $^{31}$P concentration of $5\times 10^{17}\;\rm cm^{-3}$ at a depth of $\sim$ 16 nm from the surface, calculated using the TRIM software package.\cite{Ziegler2008} The shaded blue region denotes the $^{75}$As concentration below the metal–insulator transition threshold ($7.8 \times 10^{18}\;\rm cm^{-3}$).\cite{Newman1983} This supports the presence of an $^{75}$As donor population in the active device region.

\newpage
\section{Lock-in EDMR simulations}

\begin{figure}[h]
\begin{center}
\rotatebox{0}{\includegraphics[width=0.5\linewidth]{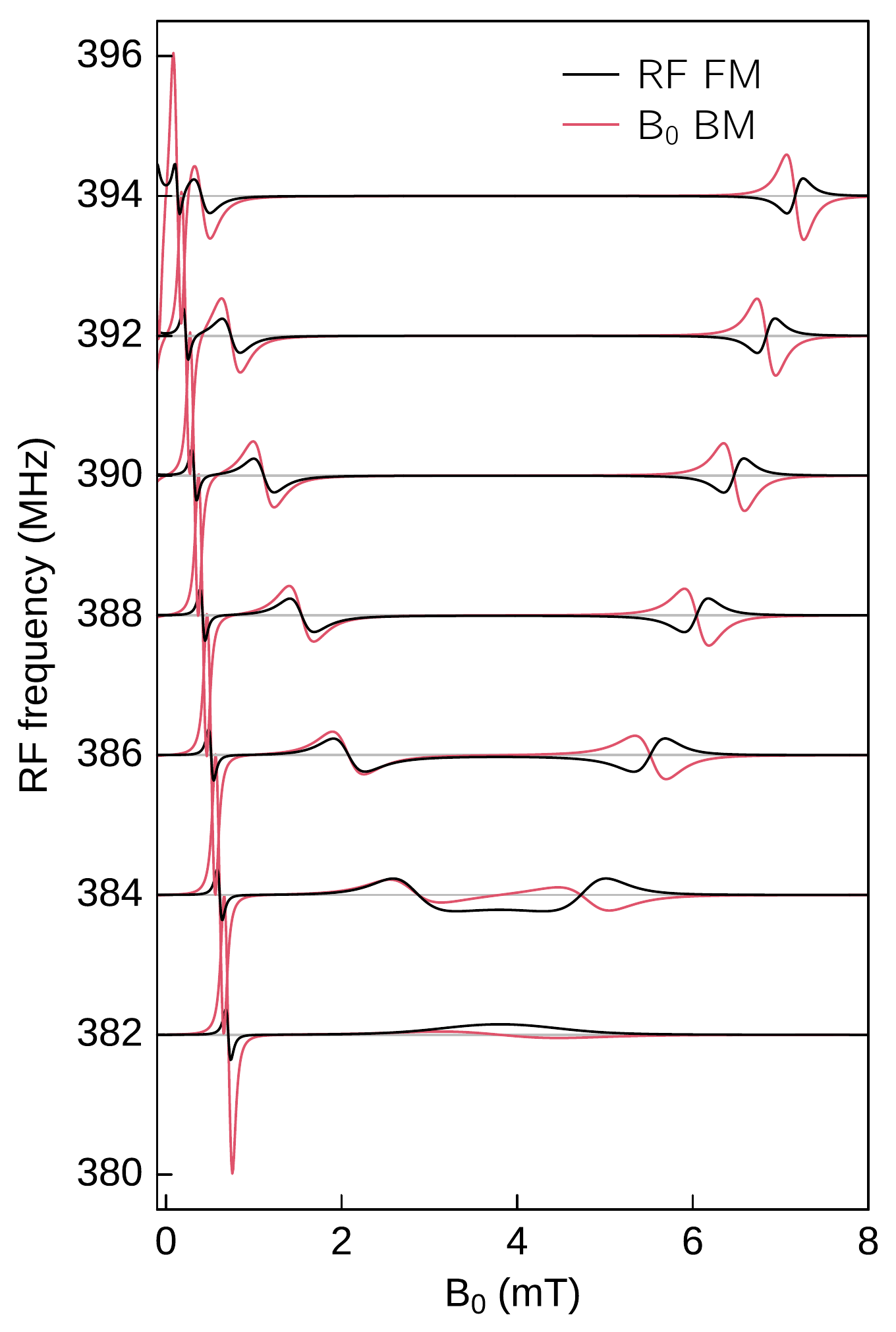}}
\end{center}
\caption{Simulated lock-in EDMR spectra near the
clock transition (CT) turning point, comparing RF frequency modulation (FM) and
magnetic field modulation (BM).}
\label{S2}
\end{figure}

RF frequency modulation (RF-FM) lock-in detection was used to measure the small EDMR signals described in the main text. This approach avoids parasitic device and PCB resonances and provides stable phase-sensitive detection across the full field range.


However, when $d f/d B_0$ varies with applied magnetic field, RF-FM leads to linewidth broadening (which may become asymmetric) and phase inversions of the lock-in signal. To illustrate these effects, EDMR spectra were simulated as a function of $B_0$ (Fig.~\ref{S2}). RF-FM is compared with conventional magnetic
field modulation (BM).

Each allowed transition was modelled  using a Lorentzian absorption profile,

\begin{equation}
L(\delta f) = \frac{1}{1+\left( \delta f/\gamma \right)^2},
\label{eq:Lorentzian}
\end{equation}

\noindent where $\delta f$ is the detuning from resonance and $\gamma$ is the half-width at half-maximum (HWHM). A value $\gamma \sim 1$~MHz was used, consistent with the experimentally observed linewidths reported in the main text.

For FM detection, the instantaneous transition frequency was taken as $f(t) = f_0 + \Delta f_{\mathrm{mod}} \sin(\omega t)$ where $\Delta f_{\mathrm{mod}}$ is the RF-FM amplitude and $\omega$ is the modulation frequency. In the field domain this corresponds to an effective modulation amplitude,

\begin{equation}
\Delta B_{\mathrm{FM}} =
\left( \frac{df}{d B_0} \right)^{-1}
\Delta f_{\mathrm{mod}},
\label{eq:BFM}
\end{equation}

\noindent which inherits both the magnitude and sign of $\left(d f/d B_0\right)^{-1}$.

For BM detection, the applied modulation is directly $B_0(t) = B_0 + \Delta B_{\mathrm{mod}} \sin(\omega t)$, independent of the spectroscopic dispersion.

The lock-in detected response was obtained by evaluating the time-dependent Lorentzian signal, multiplying by the reference modulation, and averaging over one modulation period. This procedure produces the first harmonic derivative signal measured experimentally.

As shown in Fig.~\ref{S2}, FM and BM exhibit qualitatively different behaviour when sweeping $B_0$. In BM, the modulation amplitude in field space remains constant. In contrast, when $d f/d B_0$ changes sign across the clock transition turning point, the effective modulation reverses phase, leading to a corresponding inversion of the lock-in signal.

The total EDMR signal was constructed by summing the contributions of all allowed transitions obtained from diagonalization of the donor Hamiltonian as described in the main text. These were weighted by the transition strengths. 

This simplified approach captures the field dependence of transition strengths, the distortion of lineshapes near the clock transition turning point, and the enhanced sensitivity to hyperfine and modulation-induced broadening as $d f/d B_0 \rightarrow 0$.

\newpage
\section{Line-width model fits}

\begin{figure}[h]
\begin{center}
\rotatebox{0}{\includegraphics[width=0.5\linewidth]{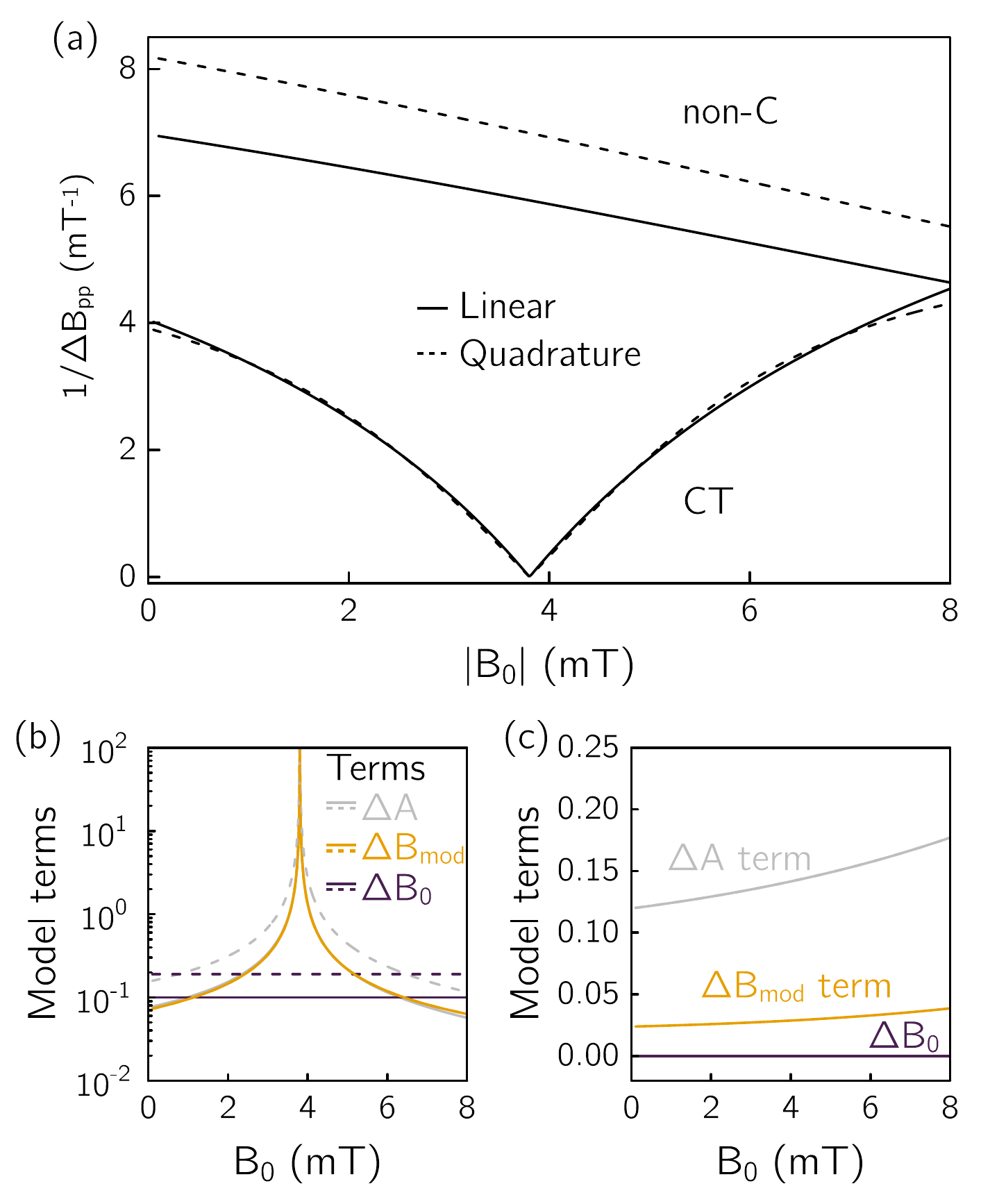}}
\end{center}
\caption{(a) Inverse of the peak-to-peak linewidth, $1/\Delta B_{\mathrm{pp}}$ from Eq.~\ref{eq:linewidth} (solid lines) and from a model with the terms added in quadrature (dashed lines) for the CT pair and non-C transitions. Model terms from (a) for (b) the CT pair and (c) the non-C  transitions.}
\label{S3}
\end{figure}

Fig.~\ref{S3}(a) shows the results of the fits to the data in  Fig.~4 of the main text, which were based on the equation:

\begin{equation}
\Delta B_{\mathrm{pp}} = 
\left( \Delta B_0 \right) +
\left( \frac{dB_0}{df} \right)
\left( \frac{df}{dA} \Delta A \right) +
\left( \Delta B_{\mathrm{mod}} \right).
\label{eq:linewidth}
\end{equation}


\noindent This equation is a linear sum of magnetic ($\Delta B_0$), hyperfine ($\Delta A$) and modulation broadening ($\Delta B_{\mathrm{mod}}$) contributions. Also shown are the results of fits to the same data using a model which adds these three terms in quadrature (root-sum-square, dashed line in Fig.~\ref{S3}(a)). The linear and quadrature models correspond to different assumptions about how the sources of disorder are correlated. The linear equation assumes that the various terms are not statistically independent and may arise from correlated broadening mechanisms. The quadrature model assumes independent sources. For the CT pair, both models provide comparable fits to the data within experimental uncertainty.

The model was also applied to the non-C transition data (indicated in Fig.~\ref{S3}(a)). Values of $\Delta B_0$ were set to zero which resulted in a vertical offset. These fits do not capture the rapid increase in broadening of the non-C transition which suggests an additional field-dependent broadening mechanism is operative. Since $df/dB_0$ is larger for this transition it may be more sensitive to magnetic noise. Changes in relaxation times at this transition may also impact the power broadening.

Each of the three terms in Eq.~\ref{eq:linewidth} and the corresponding quadrature version for the CT pair and non-C transitions are plotted separately in Fig.~\ref{S3}(b) and (c), respectively. This shows their relative contributions to the resonance linewidths as a function of applied magnetic field. The $\Delta A$ term is the strongest in the vicinity of the CT turning point (3.8~mT). For the non-C transition, $\Delta A$ is also the dominant contribution to the linewidth model across the entire spectral range investigated.

\section{Donor-P$_{b0}$ EDMR}
\begin{figure}[h]
\begin{center}
\rotatebox{0}{\includegraphics[width=0.5\linewidth]{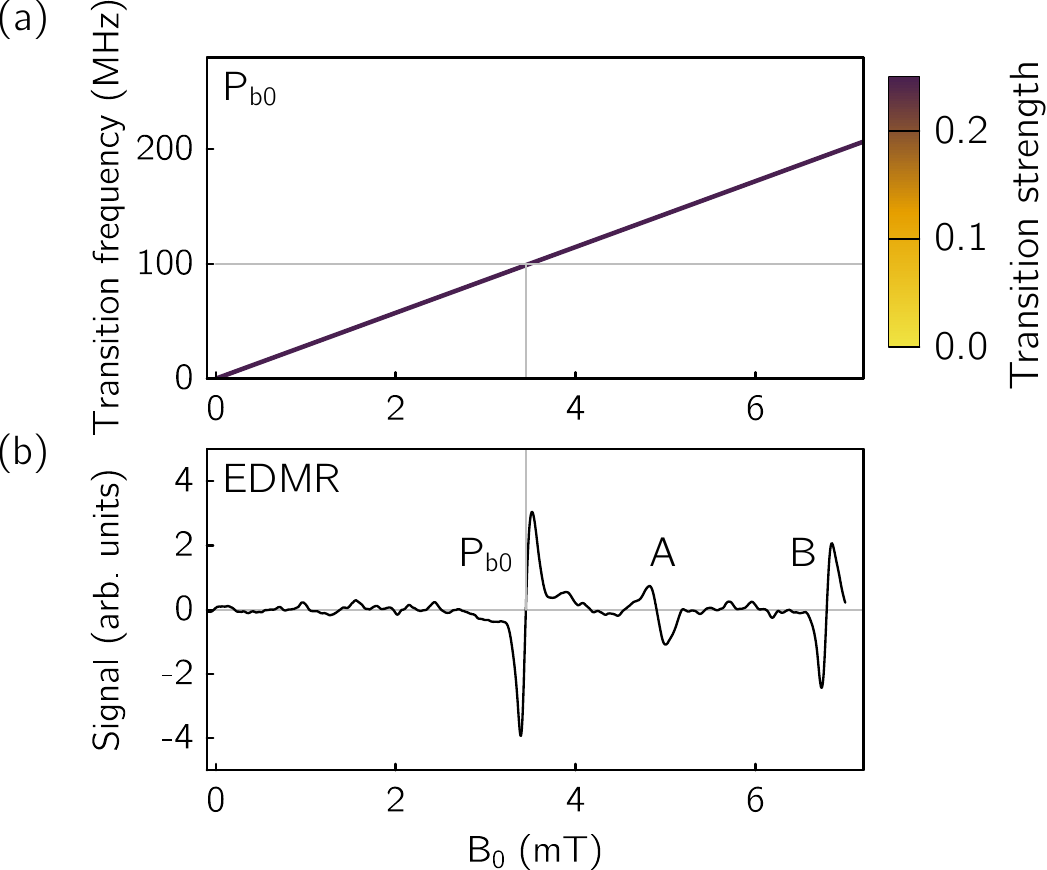}}
\end{center}
\caption{(a) Calculated low magnetic field transition energy for the P$_{b0}$. The strength of the transition is encoded in the colour scale shown. (b) EDMR spectrum recorded at 100~MHz from the same sample as that studied in the main text. The gray vertical line denotes the position of the P$_{b0}$ resonance in the spectrum. The two other peaks labelled A and B are potentially associated with higher order harmonic resonances from ${}^{75}$As and P$_{b0}$ spins, respectively.}
\label{S4}
\end{figure}

Fig. ~\ref{S4}(a)  shows the expected transition frequency versus applied magnetic field for the dangling bond centre, $P_{b0}$ with $g \approx 2$. The $P_{b0}$ centres reside at the Si-SiO$_2$ interface and are involved in the spin-dependent recombination process as discussed in the main body of text. Fig.~\ref{S4}(b) is an EDMR spectrum collected at an RF frequency of 100 MHz showing the $P_{b0}$ resonance at $B \approx 3.5$ mT, confirming their presence in the sample.
Two additional features labelled “A”, at $B \approx 4.9$ mT, and “B”, at $ B \approx 6.8$ mT, potentially arise from RF harmonics generated at high drive powers with “A” corresponding to a $^{75}$As resonance at 300 MHz and “B” to a P$_{b0}$ resonance at 200 MHz.
\newpage
\bibliography{bib}
